\documentclass{appolb}

\begin{document}
\pagestyle{plain}
\newcount\eLiNe\eLiNe=\inputlineno\advance\eLiNe by -1
\title{{Fast Timing Detectors for Forward Protons at the LHC}
}
\author{Michael ALBROW
\address{Fermi National Accelerator Laboratory, Batavia, IL 60510, USA}}
\maketitle

\begin{abstract}
I discuss the development of high precision timing detectors for high momentum
protons at the LHC, and their application in studying exclusive Higgs boson
production.

\end{abstract}

Measurements of the time of flight (ToF) of particles in high energy physics are usually 
used to measure the speed of particles between two known space-points and, together with 
the energy or momentum, to determine their mass and hence identity 
(usually $\pi^\pm$, K$^\pm$ or $p/\bar{p}$). Less common is its use to determine the position of 
origin of an ultra-relativistic particle, or a photon. 
When two oppositely directed particles are timed relative to each other, with a time difference $\Delta t$, one can 
determine the point of origin $x$ with a precision $\Delta x =
\frac{1}{\sqrt{2}}c.\Delta t$, 
(c = 3mm/10ps), 
if they came from the same point, or determine whether they did indeed have a common 
origin. ToF between two $\gamma$-rays can be used for localization of radioactive decays 
in positron emission tomography, a potentially important application still in the 
development stage. Precision $\Delta t$ measurements between  two 
outgoing protons can greatly reduce backgrounds to the exclusive reaction $p + p \rightarrow p + 
X + p$ 
at the LHC, where $X$ may be a Higgs boson, coming from cases where the outgoing protons do not come from the same
$pp$ collision
(pile-up), as first proposed in Ref.\cite{aandr}. 
The protons travel in opposite directions along the beam pipes together with the proton bunches until they are deflected 
sideways (because of their lower momemtum) by LHC magnets, where they can be detected in small 
($\approx 1$cm$^2$) detectors. Precision track measurements ($\sim 5 \mu $m over 8m) 
determine their momentum loss, 
and hence the mass of the produced state $X$.
    Groups in both CMS and ATLAS are pursuing R\&D leading hopefully to approval of these 
    very forward (at 240m and eventually 420m downstream) proton detectors. The physics 
    goal is to measure the properties of states such as the Higgs boson (if it exists) in a unique way. One selects very rare events (of 
    order 1 in 10$^{12}$
    collisions) with a Higgs candidate in the central detectors, both protons measured and no other
    particles produced. Measurements at the Tevatron of related processes (with $X = \chi_c, \gamma\gamma$, 
    di-Jet) 
    provide constraints on the (few fb) cross section~\cite{aandjeff}. 
    High luminosity must be used with perhaps about 20-30 intersections per bunch crossing, 
    every 25ns. Most $p + X + p$ events will be pile-up background, with the protons from 
    different collisions, and must be rejected. There are some kinematic constraints, but 
    powerful rejection comes from ToF $\Delta t$ measurements, and our goal 
    of $\sigma(\Delta t)$ = 10ps 
    localizes the collision point, iff they came from the same collision, to $\sigma(z)$ = 2.1mm, 
    while the collision region has length $L \sim$ 50mm ($\sigma$), and this can be matched to the 
    much better known H (WW etc) 
    vertex. Fortunately the area needed for proton detection is 
    only about 1cm$^2$, and the protons have energy near the beam energy of 7 TeV so v $\sim$ c, and 
    we can put thick (if needed) timing detectors after tracking. Requirements on the detectors 
    are that they be radiation hard to $\sim$ 10$^{15}$ p/cm$^2$ and they 
    must be edgeless, 
    i.e. active within $\sim$ 100$\mu$m of their edge, because we need to detect protons only 
    a few mm from the beam. The edgelessness is a strong constraint on the design. They must 
    also have electronics capable of being read out every LHC bunch crossing (for about 1ns every 25ns).
   Cherenkov light is prompt and good for timing, and we use that, either in a tube 
   of gas~\cite{kp} or in quartz (fused silica) bars. We 
   are studying two photon detectors with quartz bars, micro-channel plate 
   photomultipliers (MCP-PMTs) or silicon solid state photomultipliers (SiPMs). I will 
   describe beam tests of each type at Fermilab.
    The principle of the QUARTIC detector (QUARtz TIming Cherenkov) is that if several long 
    quartz bars are inclined (with respect to the protons) at the Cherenkov angle, $\theta_{ch} = 
    $cos$^{-1}(1/n) \approx 48^\circ$, and the MCP-PMT face is normal to the bars, 
    the light 
    generated in each bar in the direction of the PMT arrives simultaneously. Consider it 
    as a light wavefront propagating along the bars and parallel to the PMT window. The 
    MCP-PMT can be single-anode, as in the 40mm diameter PHOTEK PMT240, or each bar can 
    be coupled to a different anode pad, as in the Burle/Photonis 85011 (8 x 8 pads each 
    5mm x 5mm, with 10$\mu$m pores). We tested both types. The PMT240 anode design is isochronous; we 
    tested with a PiLas pulsed laser the response at different positions over the window 
    and found it to be the same within measurement errors ($\leq$ 2ps).  An advantage of the 
    8-bar/pad detector is that one makes 8 measurements, each with worse resolution, but 
    that allows the electronics to be less performant per channel, and one should recover a factor
    1/$\sqrt{8}$ by combining the measurements. They are not, however, independent as there 
    is both optical and electrical cross-talk between adjacent channels, but with the isochronous
    design this is not an issue.
Another advantage of the 8 x 8 pad detector is that if two protons pass through within the 
same bunch crossing they may be separately timed, but only if in separate rows and cross-talk 
is negligible. We have tested both the PMT240 with 1 - 5 bars each 5mm square, and the Photonis 
with up to 8 bars in a row. 

I discuss first the single anode PMT240 results. The bar housings were made at Fermilab by
electro-erosion from a solid block of aluminium. Channels were made for the bars, square with
rounded corners such that the bars did not touch the sides, except at the corners, to maintain
excellent total internal reflection, TIR, along the bar. A spring applied pressure at the bar-PMT
interface, with optical grease (which is probably not needed). We placed two identical
detectors A and B (a) On the same side of the
beam, with the bars horizontal. In this case the time difference $\Delta$t = t(A) - t(B) is independent of the horizontal
position of the track; the spread is only 2mm as defined by the trigger counter, but that corresponds to about
15ps time difference! In the experiment and in future tests the track position will be known 
to $\sigma(x) \sim 10 \mu$m. (b) On opposite sides, in which case the time sum should be independent of $x$ and the time
difference proportional to $x$, with $\frac{d\Delta t}{dx} = n/c$. Even with the constant fraction discriminators
(CFD) we found a residual correlation between the time difference $\Delta$t and the pulse heights, and we
applied a linear ``time-slewing" correction. Then the $\Delta$t distribution was a good fit to a Gaussian
distribution with $\sigma(\Delta$t) = 23.2ps.(The TDC was calibrated with a delay cable.) Comparison with a
third reference counter showed that counters A and B had the same resolution, 16.0$\pm$0.3ps each. The
combination A+B treated as a single detector (``double Quartic"), as it will be in the experiment, then 
has $\sigma$(A+B) = 23.2ps/2 = 11.6ps. 
We measured the dependence of the resolution on the number of
bars, from 2 - 5, and found $\sigma(t) = \frac{1}{\sqrt{N(bars)}}$ as expected, showing that the five bars
contribute equally; with 5 bars the number of photoelectrons per proton is $\sim 20-25$. 

  Using the double Quartic as a reference we made studies with a single 15cm long bar, inclined at 48$^\circ$,
  with a PHOTEK 210 (single anode, 10mm diameter) tube. Longer bars have the advantage of having the MCP
  further from the beam and hence in lower radiation. The main disadvantage is chromatic dispersion along the
  bar; the more intense blue light ($\lambda$ = 200nm) is slower than red light ($\lambda$ = 550nm) 
  by 3.0ps/cm. We measured a degradation of the time resolution $\sigma(\Delta$t) of about 1 ps per cm of extra bar
  length. The effect of chromatic dispersion can be reduced with color filters on the MCP window. With the
  proton at the far end of the 15cm bar we tried both red-pass (Edmund Optics 62-974, 
  $\Lambda$(min) = 400nm) and 
  blue-pass (EO 49-823, $\Lambda$(max) = 400nm) filters, and found the blue-pass filter ($\sim$15\%) better
  despite the lower pulse height. The red-pass filter gave about the same resolution as with no filter. 
  This test should be repeated. It has been suggested to replace the quartz bars with quartz
  fiber bundles; this could give some flexibility in mapping the detector cells to MCP anode pads (in a multi-anode
  MCP), enabling finer binning near the beam where the particle density is highest. As an initial test we tried
  a single fibre bundle with the same dimensions as the bar, and found a $\sim 30\%$ better resolution and a weaker
  dependence on bar length. This is a very preliminary and unexpected result; we did not have time to repeat it,
  but will investigate further, as well as combinations of fiber bundles with filters. A double-Quartic
  resolution better than our goal of $\sigma$ = 10ps is in reach (with adequate electronics).
  
  We also tested a detector with silica aerogel as Cherenkov radiator,with a 45$^\circ$ plane mirror
  reflecting the
  light to a PHOTEK240 MCP-PMT. Aerogel is cheap, radiation hard and has a refractive index (which can be tuned in
  production) $\sim$ 1.01-1.05 with essentially no dispersion in the visible range. It is however fragile. We used a 25mm long block with n =
  1.03, $\theta_{ch}$ = 13.9$^\circ$, and about 13 p.e. per proton. We measured a time resolution $\sigma(t)$ = 31ps.
  The amount of material in the beam is very small and one could have several such detectors in line to get good
  resolution. However the light cannot be focussed to a small area, and the large tubes are expensive and close to the
  beam. One might develop the aerogel idea, e.g. replacing the MCP-PMT with an array of SiPMs. 

We tested arrays of SiPMs with quartz bars in front (Q-SiPMs), in line with the beam. For a particle parallel to the axis of a
parallel sided bar, all the Cherenkov light is totally internally reflected to the back, for any refractive index, as the Cherenkov angle and TIR
angle are complementary (cos $\theta_{Ch} = \frac{1}{n} =$ sin $\theta_{TIR}$). 
We used SiPMs made by STM Catania,IT (with
thanks), area 3.5$\times$3.5mm$^2$, with 4900 cells of 50$\times 50\mu$m$^2$; each cell is an avalanche photodiode. We
varied the length of quartz radiator from 5mm to 30mm and found a continuous improvement 
 to $\sigma \sim$ 15ps at 30mm (even though the time difference
between light arriving from the front and back is then about 120ps). More light helps, even if
it is not prompt. In our application we can have many (e.g. 10 - 20) such detectors in line, and fit the many time
measurements to a line (``time-track") similar to multi-layer track chambers. This provides continuous measurements of the
resolution, efficiency, and calibration of each detector. Each layer can consist of an array of many individual detectors to provide
multihit capability. We tested six Q-SiPMs in-line with 10mm Q-bars, through an ORTEC VT120 $\times$20 fast pre-amp into a DRS4
``scope-guts'' waveform recorder (200ps/sampling). Simply making a linear fit from 10\%-90\% of the amplitude and
extrapolating to zero amplitude was close to optimal, and we found $\sigma(t) \sim$ 30ps per detector.  
An issue is the lifetime of the SiPMs close to the LHC beam, however they could be made for simple
replacement every year or so if needed, as they are relatively inexpensive. Alternatively the SiPMs can be at the end of
48$^\circ$ bars as in the \textsc{quartic}, replacing the MCP-PMT. We will test this configuration in 2011.
An attractive possibility is to make SiPMs with a ``mini-strip" geometry, e.g. 1mm
individually read-out strips,
and use a quartz fiber bundle bar at 48$^\circ$ to the beam. This could give independent timing on particles only 2mm 
apart in $x$, and keeps the SiPMs far enough from the beams to minimize radiation damage. With multiple layers,
we would emulate track chambers but in the time domain.

 The Q-SiPM combination makes a very compact fast timing counter which could be used as a (directional) monitor of particle fluxes just outside
 the LHC beam pipe, bunch-by-bunch, and with time resolution less than the proton bunch length. The LHC Instrumentation group are looking into this possibility.
 
 An interesting possibility is to develop a ``GHz streak camera" using a silicon pixel detector. Photoelectrons from a photocathode are accelerated and focused to a small spot, which can be swept
 with x-y sweeping electrodes
 over the pixel detector with the LHC 400 MHz (or other) RF into a circle, converting time into spatial position. 
 I have not developed this idea, but Amur Margaryan at this meeting described a similar project.
 
 Finally I mention two more ideas for rejecting pile-up background in the $p+H+p$ search, which could be done with large
 area (many m$^2$) pixellated timing detectors. Covering forward discs with such detectors to time each
 charged particle arriving there, enables the time of the collision at their origin to be reconstructed. Those events
 are \emph{background}, as the $p+H+p$ events do not usually have such forward particles. Even more ambitious, is to
 cover the central region with good timing detectors to make precision time measurements of all collisions, to
 correlate with the time from the protons. If thin enough, they might be close to the collision region, with fine
 enough granularity to time all tracks. Simply (!) add fast timing capability to silicon tracking detectors! Both these
 schemes could provide additional pile-up rejection.
 
  Some of these results have been published~\cite{ronzhin}.
  I especially thank S.Malik, E.Ramberg, A.Ronzhin, A.Zatserklyaniy for close collaboration on the tests.

\end{document}